\documentclass[12pt]{article}
\usepackage[dvips]{graphicx}   

\begin{document}
\title{The simple complex numbers}
\author{Jaros\a{l}aw Zale\'sny  \\  \\
\emph{Institute of Physics, Szczecin University of Technology,}   \\
\emph{Al. Piast\'ow 48, 70-310 Szczecin, Poland}}
\date{}
\maketitle

\begin{abstract}
A new simple geometrical interpretation of complex numbers is presented. It differs from their usual interpretation as points in the complex plane. From the new point of view  the 'complex numbers' are rather operations on vectors than 'points'. Moreover, in this approach the real, imaginary and complex numbers have similar interpretation. They are simply some operations on vectors. The presented interpretation is simpler, more natural, and  better adjusted to possible applications in geometry and physics than the usual one, especially for describing rotations in a plane. The relation of the new approach to the usual interpretation and especially to the notion of 'complex plane' is also clarified in the paper. The new interpretation of 'complex numbers' gives new insight into their applications in physics, which is demonstrated by some elementary examples in mechanics and optics.  
\\

PACS numbers: 01.40.-d, 01.70.+w, 01.90.+g, 02.90.+p
\end{abstract}

\section{Introduction}
\label{sec: 1}

I believe that in some wider sense mathematics is a part of physics, although I admit that  rather a small number of mathematicians share this view. Let me explain my point of view with the example of natural numbers. Mathematicians like to construct the numbers on the basis of the set theory as so called cardinal numbers. 'All we need' they say 'is the notion of the empty set.' It is the only 'object' allowing to construct all the natural numbers. A physicist feels that some important features escape mathematicians' attention. From the physical point of view it is obvious that the notion of natural numbers appeared in the human mind because we could distinguish in 'reality' some objects, like stones or cows, which are separate in space and time. Moreover, even mathematicians, when they try to express the successive natural numbers in terms of set theory, they have to write more and more 'separate objects' - 'circles' $\emptyset$ (symbols of the empty set). The notion of natural numbers appears in the process which mathematicians name abstraction or generalization. We tend to forget that in fact the objects are stones, cows or circles and what remains are numbers. A number is some characteristic of the set of objects, like color or smell. In fact, notions are the only 'tools' used by the human mind to explore the surrounding world. We believe that beyond our minds there exists the Universe, nevertheless the 'reality' is not simply given to us. In fact the human world consists exclusively of conscious and unconscious, innate (senses) and learned (ideas) notions. All things are merely  abstractions. In this sense all in the human world is some kind of mathematics. Nevertheless, the systems of notions have been formed in the process of biological or cultural evolution in the course of the contact of the mind with the Universe. Therefore these notions reveal a piece of truth about  'reality'. Consequently it is nothing peculiar in the fact that mathematics is so well suited to the 'physical world'. It suits itself.

Physicists are people who are interested in describing motions in space and time. What kind of an abstract 'object' would be the most suitable tool to reach this goal? I claim that it is the notion of a vector. The prototype of a vector is a displacement. One may also think of it as of an oriented segment of line. The main purpose of this paper is to examine operations on vectors. This will lead to a very natural, almost obvious interpretation of complex numbers. I wish I was taught at school about complex numbers in this way. Unfortunately, one may look into any mathematical textbook to see that something 'mysterious' still remains in the teaching about complex numbers, e.g. the imaginary unit often is defined as $i=\sqrt{-1}$, though every student knows that such a square root of $-1$ does not exist. To clarify the matter I have written this paper. I hope it will be interesting not only for undergraduate students but also for all the people who use complex numbers in every day practice.

\section{Operations in a plane}
\label{sec: 2}

Consider the arithmetical expression $2\cdot 3$. Do both numbers in this expression have the same meaning? We used to say 'two times three'. Let $3$ have the meaning of 'three apples'. What is then the result of the above expression? Certainly 'six apples'. Therefore $2$ cannot mean  'two apples', because multiplication 'apples by apples' does not make sense. $2$ means here an operation. It  means 'two times'. Equivalently, we may say that $2$ is here an operator acting on objects (the apples). In order to distinguish operators from objects, we may use such a notation as $\hat{2}\cdot 3$. Note that the fact that $\hat{3}\cdot2$ is also 'six apples' needs some justification or may be taken as a 'law of Nature'. So, arithmetical expressions have the following structure: $operator\to objects$. But things are not always so simple. Consider the expression $2\cdot3\cdot5$. Suppose  $5$ means now 'five apples'. Certainly the full expression means $30$ apples. This means that first $5$ apples have been taken 'three times' and next the result of this operation has then been taken 'two times'. Therefore this time $2\cdot3$ has the meaning of a product of operations. Thus the full expression could be written more properly as $\hat{2}\cdot\hat{3}\cdot5$. Generally any arithmetical expression has the form: $operator1\cdots operatorN\to objects$. The counted  'apples' are treated as identical. Actually, you do not count real apples, but rather some abstract objects. Note  however that e.g. 'three apples'  could be rewritten as $3=\hat{3}\cdot 1$. Therefore, in fact we need only \emph{one abstract} object ('an apple'). By performing operations on it we can 'create' as many 'apples' as we want. Moreover, we see that multiplication means a product of operations and not product of objects. Thus the expression $2\cdot 3$ should be properly understood as $\hat{2}\cdot \hat{3}\cdot 1$. Now the numbers $2$ and $3$ have the same meaning as operations on the abstract object $1$.
Certainly, there is no need to use integers as the operations. They could be real numbers as well. 

Consider now an algebraic equation, e.g. $2 x^{2}+3 x + 4 = 0$. What is the meaning of $x$? An object or an operation? If one insists on  treating $x$ as an object (e.g. apples) then $x$ and the numbers $2, 3, 4$ have to possess (different) physical units.  It is better, however, to understand the equation in the following way: $\{\hat{2} \hat{x}^{2}+\hat{3} \hat{x} + \hat{4}\}{~} object = \hat{0}{~}object$. Then all quantities in it are pure numbers, except of course of an $object$. It contains the 'physics' of the problem. Thus, generally, algebraic equations should be  understood rather as operation equations even when the 'hats' over the operators and the symbol of $object$ are omitted as is usually the case.

From a mathematical point of view, the most interesting feature of a collection of 'apples' is their number. However, there are many cases when not only 'number' but also 'direction of the number' is important. This leads us to geometry. Therefore, from now on we will treat \emph{vectors} as abstract objects. 
Certainly, we can add vectors. Addition is a very natural activity. It is achieved by placing the initial point of the next vector on the final point of the previous vector ('tip to tail' method). The resultant vector (the sum) means straightforward displacement from the initial point of the first vector to the final point of the last vector. The prototype of any vector is a displacement. One can also treat a vector as the following recipe or instruction \emph{'go a certain number of steps in a given direction'}. E.g. the sum of two instructions $\mathbf{v}_{1}=$ \emph{'go three steps north'} and $\mathbf{v}_{2}=$\emph{'go four steps east'} can be replaced with the one $\mathbf{v}_{3}=$\emph{'go five steps (almost) north east'}, see \mbox{Figure 1.} The instructions may be treated as 'free vectors' so long as initial points of displacements are not specified.
\begin{figure}[!h]
\begin{center}
\includegraphics[width=0.4\textwidth]{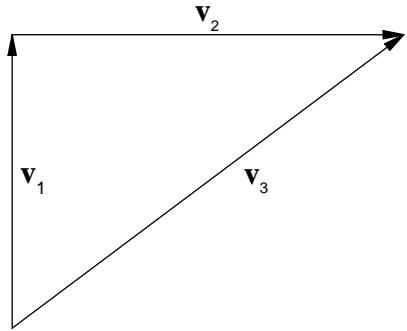}
\label{fig:FiguresComplex_gr1}
\caption{An example illustrating vector addition, $\mathbf{v}_{1}+\mathbf{v}_{2}=\mathbf{v}_{3}$}
\end{center}
\end{figure}

Consider any vector $\mathbf{v}$.  One of the simplest operations on a vector is to shorten or lengthen it, i.e., to multiply the vector by  a positive real number $a$. This operation yields another vector $\mathbf{v'}$, see Figure 2. The number is an operator, but from now on I will omit the 'hats' over the operators.  The operation can be written as
\begin{equation}
\label{1}
\mathbf{v'}= a \mathbf{v}
\end{equation}
In fact, the equation (\ref{1}) can be considered not only for positive but for any real number $a$. For a positive $a$ the vectors $\mathbf{v'}$ and $\mathbf{v}$ point in the same direction. $a=0$ defines the null vector. A negative  $a$ reverses direction of the vector $\mathbf{v}$ and changes its length $\vert a \vert$ times. The meaning of the negative sign of a vector has a natural geometrical explanation. Starting from any point $P$ according to instruction $\mathbf{v}$ and then going back according to $-\mathbf{v}$ one returns again to $P$. 
That is, any $\mathbf{v}$ and $-\mathbf{v}$ are mutually opposite vectors in that sense that $\mathbf{v} + (-\mathbf{v})\equiv\mathbf{v}-\mathbf{v}\equiv(1-1)\mathbf{v}=0{~}\mathbf{v}=\mathbf{0}$. Thus, $1$ and $-1$ are reverse operations. For $\vert a \vert \neq 1$ the magnitudes of the vectors are different. Note that we are not forced to know the absolute magnitudes of the vectors. If we choose \emph{any} vector as a 'unit vector', e.g. vector $\mathbf{v}$, then the operation (\ref{1}) defines a dimensionless measure of any other vector as $\vert a \vert$.  Equation (\ref{1}) makes it also possible to assign to any operation $a$ a point on the straight line, i.e., to construct the numerical axis. The construction certainly depends on the choice of the 'unit vector'. In fact, the 'unit vector' is the only abstract object we need. All other vectors parallel to it can be obtained with the help of operation (\ref{1}). In that way we have achieved the interpretation of real numbers as operators acting on vectors. Multiplication of real numbers, e.g., $a, b, c$,  is in fact a product of operations of the type (\ref{1}), i.e., $a{~}b{~}c{~}\mathbf{v}$. In this interpretation the rules of signs like 'minus times minus is plus' and so on, are obvious and natural from a geometrical point of view. 
\begin{figure}[!h]
\begin{center}
\includegraphics[width=0.4\textwidth]{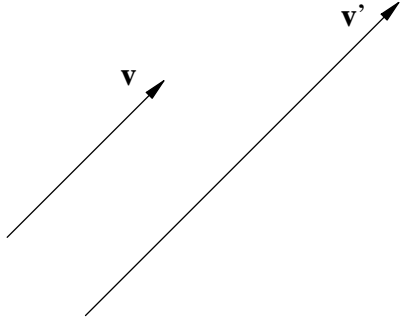}
\label{fig:FiguresComplex_gr2}
\caption{The operation of multiplying a vector by a number, $\mathbf{v'}= 2 \mathbf{v}$}
\end{center}
\end{figure}

The operation (\ref{1}) is unsophisticated. All the vectors produced by it are parallel. Now we want to omit this restriction. Suppose that we want to construct \emph{any} vector in a \emph{plane}. To this end it is enough to introduce only one more operation. Every soldier knows the command 'left turn!'. It is simply 90-degree left turn. We will denote this operation by the letter $i$. The operator $i$ simply rotates a given vector $\mathbf{v}$ by a 90-degree angle in the plane. 
\begin{equation}
\label{2}
\mathbf{v'}= i \mathbf{v}
\end{equation}
Thus, the vectors $\mathbf{v'}$ and $\mathbf{v}$ are perpendicular. Note that two commands 'left turn!' produce 'about turn!', that is a 180-degree turn, see Figure 3.  It can be written as
\begin{equation}
\label{3}
i i \mathbf{v} \equiv i^{2} \mathbf{v}= - \mathbf{v}
\end{equation}
or as an operator equation 
\begin{equation}
\label{4}
i^{2} = - 1
\end{equation}
For historical reasons the operator $i$ is named the 'imaginary unit'. However this name is confusing. The geometrical meaning of the operator is obvious and natural. The confusion has its roots in that people often ignore the fact that algebraic equations should be properly treated  as operation equations.  E.g., solutions of the equation $(x^{2}+1) \mathbf{v}=0{~}\mathbf{v}$ are operations $i$ or $-i$, and solutions of the equation $(x^{2}-1) \mathbf{v}=0{~}\mathbf{v}$ are operations $1$ or $-1$. And there is nothing strange in neither cases. Both have clear geometrical meanings.
\begin{figure}[!h]
\begin{center}
\includegraphics[width=0.4\textwidth]{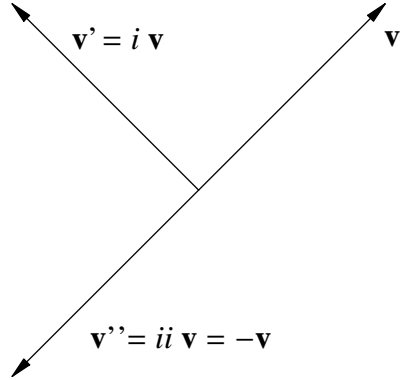}
\label{fig:FiguresComplex_gr3}
\caption{The operation $i$}
\end{center}
\end{figure}

Suppose now that there are two arbitrary vectors in the plane, e.g., $\mathbf{v}$ and $\mathbf{v'}$, see Figure 4. We want to find an operation which converts vector $\mathbf{v}$ to vector $\mathbf{v'}$. To this end we can build a right-angled triangle using the two vectors. To the vector $\mathbf{v}$ multiplied by an appropriate number $a$, we add the vector $i\mathbf{v}$ multiplied by appropriate number $b$. In this way we obtain vector $\mathbf{v'}$. Note that the operation of multiplication by number and the operation $i$ commute. The result can be also written in the following form
\begin{equation}
\label{5}
\mathbf{v'}=a\mathbf{v}+ i b \mathbf{v} = (a + i b)\mathbf{v} = z \mathbf{v}
\end{equation}
where the operation $z$ has been written as a sum of operations
\begin{equation}
\label{6}
z = a+ib.
\end{equation}
Traditionally $z$ is called a 'complex number'. Nevertheless, in our interpretation, it is the operation mapping a vector $\mathbf{v}$ on a vector $\mathbf{v'}$.
\begin{figure}[!h]
\begin{center}
\includegraphics[width=0.6\textwidth]{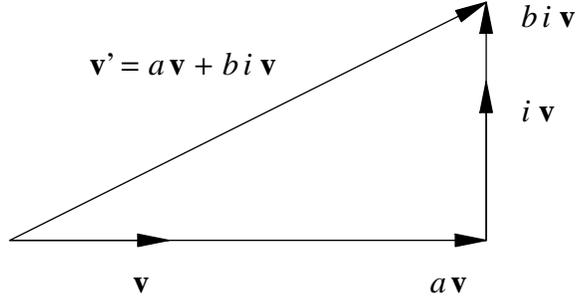}
\label{fig:FiguresComplex_gr4}
\caption{The operation $z= a+bi$}
\end{center}
\end{figure}

With the help of operation (\ref{6}) it is easy to describe the rotation of a vector in the plane. Indeed, if we want to turn the  vector $\mathbf{v}$ through some angle $\alpha$, we should take $a=\cos\alpha$ and $b=\sin\alpha$. Thus, the operation 
\begin{equation}
\label{7}
\mathbf{v'}=\cos\alpha{~} \mathbf{v}+ i \sin\alpha{~} \mathbf{v} = (\cos\alpha + i \sin\alpha)\mathbf{v} \equiv  \mathrm{e}(\alpha) \mathbf{v}
\end{equation}
describes the rotation. $\mathrm{e}(\alpha)$ is the rotation operator. Suppose that after the first rotation we make another rotation. This time rotation of the vector $\mathbf{v'}$ by an angle $\beta$
\begin{equation}
\label{8}
\mathbf{v''}= (\cos\beta + i \sin\beta)\mathbf{v'} \equiv  \mathrm{e}(\beta) \mathbf{v'}.
\end{equation}
Certainly, we can achieve the same vector $\mathbf{v''}$ turning the vector $\mathbf{v}$ through the angle $\alpha + \beta$ 
\begin{equation}
\label{9}
\mathbf{v''}= \left(\cos(\alpha+\beta) + i \sin(\alpha+\beta)\right)\mathbf{v} \equiv  \mathrm{e}(\alpha+\beta) \mathbf{v}
\end{equation}
Therefore the function $\mathrm{e}(\alpha)$ has the property
\begin{equation}
\label{10}
\mathrm{e}(\alpha+\beta) = \mathrm{e}(\alpha) {~} \mathrm{e}(\beta)
\end{equation}
Note that the order of the operators $\mathrm{e}(\alpha)$, $\mathrm{e}(\beta)$ is not important, because rotations in a plane commute, as is clear from geometrical considerations.
The property (\ref{10}) is a characteristic property of the exponential function. Thus, algebraically the rotation operator $\mathrm{e}(\alpha)$ behaves like the exponential function and can be written as $\exp{(u\alpha)}$, where $u$ is  an unknown that can be determined by considering the derivative of the operator $\mathrm{e}(\alpha)$ with respect to $\alpha$
\begin{equation}
\label{11}
\frac{d}{d\alpha}{~}\mathrm{e}(\alpha) =   \frac{d}{d\alpha}{~}(\cos\alpha + i \sin\alpha) = i (\cos\alpha + i \sin\alpha) = i \mathrm{e}(\alpha)
\end{equation}
In this way the rotation operator takes the form
\begin{equation}
\label{12}
\mathrm{e}(\alpha) =   e^{i \alpha}
\end{equation}
and the operation (\ref{7}) can be rewritten in the following important
form
\begin{equation}
\label{13}
\mathbf{v'}= e^{i\alpha} \mathbf{v}
\end{equation}
This is perhaps the simplest recipe for the rotation of a vector.

In general, for any two vectors $\mathbf{v}$ and $\mathbf{v'}$ in a plane we can turn the first vector $\mathbf{v}$ in the  direction parallel to the second vector $\mathbf{v'}$ and next  change its length by multiplying with an appropriate number $r$ to obtain a vector identical with $\mathbf{v'}$. Thus, the general form for a vector transformation is    
\begin{equation}
\label{14}
\mathbf{v'}= r e^{i\alpha} \mathbf{v}
\end{equation}
A nice feature of the presented formalism is that it is coordinate-free. However, it works in a coordinate frame as well. In a Cartesian frame the vector components can be taken as follows $\mathbf{v}=(v_{x}, v_{y})$ and $\mathbf{v'}=(v'_{x}, v'_{y})$. Equation (\ref{14}) then takes the form
\begin{eqnarray}
\label{15}
v'_{x}\mathbf{\hat{x}} + v'_{y}\mathbf{\hat{y}}= r e^{i\alpha} (v_{x}\mathbf{\hat{x}} + v_{y}\mathbf{\hat{y}})=
r (\cos\alpha + i \sin\alpha) (v_{x}\mathbf{\hat{x}} + v_{y}\mathbf{\hat{y}}) = \nonumber \\
=(rv_{x}\cos\alpha  - rv_{y}\sin\alpha )  \mathbf{\hat{x}} + (rv_{y}\cos\alpha  + rv_{x}\sin\alpha )  \mathbf{\hat{y}}
\end{eqnarray}
where $\mathbf{\hat{x}}$ and $\mathbf{\hat{y}}$ are the unit vectors of the coordinate frame, and certainly $i \mathbf{\hat{x}}=\mathbf{\hat{y}}$.
This gives us automatically the transformation formulas for the vector components.

In the standard approach the 'complex numbers' are identified with points in the 'complex plane'. Let us explain now the relationship between the standard picture and the one presented here. Suppose that in the usual plane one chooses a certain Cartesian coordinate system definite by the unit vectors $\mathbf{\hat{x}}$, $\mathbf{\hat{y}}$. The points in the plane can be identified with the position vectors $\mathbf{r}=x \mathbf{\hat{x}} + y \mathbf{\hat{y}}$. 
However, all these vectors can be obtained by the use of operation $z$ (see (\ref{6})) on the unit vector $\mathbf{\hat{x}}$.  
\begin{equation}
\label{16}
z \mathbf{\hat{x}}= (x + i y)\mathbf{\hat{x}}=x \mathbf{\hat{x}}+y \mathbf{\hat{y}}=\mathbf{r} 
\end{equation}
In this way every operator $z$ may be identified with a position vector $\mathbf{r}$, i.e., with a point in the plane. 
\begin{equation}
\label{17}
z \Leftrightarrow \mathbf{r} 
\end{equation}
This is nothing but the 'complex plane'. Note however that the construction depends on the choice of the coordinate frame.
In order to avoid this problem, the standard interpretation usually claims that the 'complex plane' is some 'abstract' plane. In our approach this is unnecessary.

At the end of this section let me raise a technical question. Consider the scalar product of the vector $i\mathbf{v}$ with itself. In the standard approach one directly claims that $i\mathbf{v\cdot}i\mathbf{v}=-\mathbf{v\cdot v}=- v^{2}$, i.e., one takes out $i$ in front of the scalar product. However in our approach $i\mathbf{v}$ means a left 90-degree rotation of vector $\mathbf{v}$. Therefore the result of the scalar product is $i\mathbf{v\cdot}i\mathbf{v}=\mathbf{v\cdot v}= v^{2}$. It should be remembered that $i$ means an operation on a vector. It is not allowed to isolate the operators from vectors in this case.  In the standard approach,  vectors like $\mathbf{v}$ in the above example, are treated as some parameters only. Consequently, the vector space remains undefined. 

\section{Elementary examples in physics}
\label{sec: 3}

One of the simplest and natural applications of the presented interpretation of the 'complex numbers' is to use them for description of  circular motion in a plane. This application is based on the formula (\ref{13}). Suppose that the motion is observed from some distant point $O$. Let $\mathbf{R}$ be the position vector of an object in its circular motion. The vector may be regarded as a sum of two vectors $\mathbf{Q}$ and $\mathbf{r}$. The initial point of the vector $\mathbf{Q}$ is at the point $O$ and the final point at the center of the circle. Therefore $\mathbf{r}$ is the position vector of the object relative to the center of the circle. Suppose, for simplicity, that the center of the circle does not move. Then the vector $\mathbf{Q}$ is constant and only the direction, but not the length, of the  vector $\mathbf{r}$ depends on time $t$. The motion may then be described by the equation
\begin{equation}
\label{18}
\mathbf{R}(t)= \mathbf{Q} + \mathbf{r}(t) = \mathbf{Q} + e^{i\alpha(t)} \mathbf{r}_{0}
\end{equation}
Consider uniform circular motion, then $\alpha(t)=\omega t$, and $\omega$ is a positive constant angular velocity. 
Note that such a choice of sign means anticlockwise motion.
Taking the first and second time derivatives of the formula (\ref{18}), the velocity $\mathbf{v}$ and the centripetal acceleration $\mathbf{a}$ can  easily be found
\begin{equation}
\label{19}
\mathbf{v}(t) =   i\omega e^{i\omega t} \mathbf{r}_{0} =  i\omega{~} \mathbf{r}(t), \qquad \mathbf{a}(t) =   - \omega^{2} \mathbf{r}(t)
\end{equation}
The velocity is by no means 'imaginary' as in the  sense of the standard interpretation. Instead the vector $i \mathbf{r}(t)$ is always  perpendicular (left turn) to the vector $\mathbf{r}(t)$, which means that the velocity $\mathbf{v}(t)$ is at every moment perpendicular to the radius vector $\mathbf{r}(t)$. The relation between their magnitudes is $v=\omega r$. Note that the result has been expressed without using the notion of vector angular velocity and  the vector product. This is some advantage, since, as it is well known, the angular velocity is \emph{not} a true vector. It is rather a disguised bivector. The construction of vector product is somewhat artificial and is possible only in three-dimensional space. Nevertheless, in any-dimensional space a rotation is properly determined in a plane and not around an axis.

Another similar example refers to the description of motion in a rotating frame, e.g., on a rotating disk. Suppose that the origins of an inertial frame and a noninertial frame coincide with the center of the disk. Let at some moment $t$ the position of a 'body' be given in the inertial frame by the vector $\mathbf{r}(t)$ and let the disk rotate to the right with regard to the inertial frame. In the noninertial frame, rotating together with the disk, the position of the body is given by  the vector $\mathbf{r'}(t)$. Note that even if the body is at rest with regard to the inertial frame, i.e., vector $\mathbf{r}$ is constant, the vector $\mathbf{r'}$ rotates to the left with respect to the disk. Therefore, in general
\begin{equation}
\label{20}
\mathbf{r'}(t) =e^{i\alpha(t)}\mathbf{r}(t) 
\end{equation}
For a uniform rotation $\alpha(t)=\omega t$. Though the formula (\ref{20}) is similar to (\ref{14}), its meaning is different. In formula (\ref{14}) both vectors are different. In (\ref{20}) $\mathbf{r}$ and $\mathbf{r'}$ is in fact the same vector, but considered in different frames.   The transformation formula (\ref{20}) is the shortest way to express the geometrical content of the problem. Introducing explicitly the coordinate systems, one can easily find the transformation rules for the vector components in a similar way as was done in equation (\ref{15}). 

The first derivative of equation (\ref{20}) gives
\begin{equation}
\label{21}
\mathbf{v'}(t) =e^{i\omega t}\mathbf{v}(t) + i \omega \mathbf{r'}(t) 
\end{equation}
where $\mathbf{v}\equiv \mathbf{\dot{r}}$, $\mathbf{v'}\equiv \mathbf{\dot{r}'}$ and the dot means time derivative. The result has an obvious interpretation. The first term in (\ref{21}) is nothing but the velocity vector $\mathbf{v}$ seen from the disk, turned through the angle $\omega t$. The second term means a contribution from the linear velocity of the disk at the distance $ \mathbf{r'}$ from the disk center.

The second derivative of (\ref{20}) gives a relation between the accelerations $\mathbf{a}\equiv \mathbf{\dot{v}}$ and $\mathbf{a'}\equiv \mathbf{\dot{v}'}$
\begin{equation}
\label{22}
\mathbf{a'}(t) =e^{i\omega t}\mathbf{a}(t) + 2 i \omega \mathbf{v'}(t) +\omega^{2}\mathbf{r'}(t)  
\end{equation}
The first term means the turned inertial acceleration $\mathbf{a}$, the second term is the Coriolis acceleration, and the third term means centrifugal acceleration. In the Coriolis formula, the term $i \mathbf{v'}$ means that the acceleration is always perpendicular to the velocity of the body in the noninertial frame.

The next example comes from optics or more generally from electrodynamics. For mathematical convenience real fields are often  written in complex form. For simplicity, consider an electric plane wave
\begin{equation}
\label{23}
\mathbf{E}(t) =\mathbf{E_{0}}\cos{(\mathbf{k\cdot x}-\omega t)}= \frac{1}{2}\left(\mathbf{E_{0}}e^{i(\mathbf{k\cdot x}-\omega t)} + \mathbf{E_{0}}e^{-i(\mathbf{k\cdot x}-\omega t)}\right)
\end{equation}
Is it only a mathematical trick or is a physical meaning in the complex notation? In standard interpretation it is only a convenient trick. In our approach to complex numbers, the two complex terms on the right hand side of equation (\ref{23}) possess a clear physical sense. On the basis of equation (\ref{13}) we can claim that the term      
\begin{equation}
\label{24}
e^{i(\mathbf{k\cdot x}-\omega t)}\mathbf{E_{0}}  
\end{equation}
represents right circularly polarized wave, and the 'complex conjugate'
\begin{equation}
\label{25}
e^{-i(\mathbf{k\cdot x}-\omega t)}\mathbf{E_{0}}  
\end{equation}
represents left circularly polarized wave.  It can be inferred from Maxwell equations that the plane of rotation of the electric vector $\mathbf{E_{0}}$ (i.e., the plane in which $i$ acts) is perpendicular to the vector $\mathbf{k}$.   
The superposition of the two circularly polarized waves gives as a result the linearly polarized wave (\ref{23}). In accordance with the choice $operator\rightarrow vector$, the vector $\mathbf{E_{0}}$ has been placed on the right sides of the operators in (\ref{24}) and (\ref{25}). 

And at last, let me make some remarks on quantum mechanics (QM). This is the field of physics, where complex numbers seem to be almost necessary. The 'imaginary unit' $i$ occurs directly in Schr\"odinger equation. It is remarkable that $i$ always accompanies the Planck constant $\hbar$. In QM complex numbers appear mainly in two contexts. They give the easiest way to describe interference effects, and they serve for description of spin (e.g., they occur in Pauli matrices). Certainly, usually physicists think of them in terms of the standard interpretation. Nevertheless, see as a counter-example the papers of David Hestenes, e.g., \cite{Hestenes_spin} (see also comments in 'Final remarks' below). The presented here interpretation of complex numbers reveals however, that in QM there are \emph{no vectors} (in the sense of this paper) on which $i$ or generally complex numbers could act. Maybe it is a part of interpretative troubles of QM that it contains \emph{operators} but there are no \emph{objects}. However, even if this conjecture is right, this matter goes far beyond the intended scope of this modest paper.

Other examples can be easily given, however I leave for the reader the pleasure to invent them. I hope however, that it is now clear that the presented interpretation of 'complex numbers' can be useful and sometimes sheds new light on a problem.

\section{Final remarks}
\label{sec: 5}

The standard approach to complex numbers can be found in any standard text on the subject. A non-standard approach, in some sense very close 'in a spirit' to the ideas of this paper, but technically quite different, the reader may find in excellent papers and books on \emph{geometric algebra} of David Hestenes (Arizona State University), see e.g. \cite{Hestenes_book}. His papers and at least some important chapters of his books can easily be found in internet, therefore I give here no more references but strongly recommend his web page \cite{Hestenes_net}. In Europe, an excellent work has been done, 'in the spirit of Hestenes', by Anthony Lasenby and other people from the Cambridge group (UK). I recommend their web page \cite{Lasenby_web}. An interesting introduction on a 'physical level' to geometric algebra and its application to classical electrodynamics contains also the book of Bernard Jancewicz (Wroc\a{l}aw University, Poland) \cite{Jancewicz_book}. Geometric algebra is a powerful and nice mathematical system. The main objects in it are multivectors, and the main operations on them are Clifford and Grassmann products. In geometric algebra of a plane there is a unit bivector which may be regarded as the generator of rotations in the plane. It can be identified with the 'imaginary unit' $i$. Multiplication by the unit bivector of any vector in the plane rotates the vector by a right angle. Thus, the 'final effect' of the operation is the same as in my paper. However, to obtain this result in geometric algebra the powerful 'mathematical machinery' is engaged (Clifford product, multivectors, etc.). My approach is much simpler. Jokingly, I may call my approach as a pre-geometric-algebra approach.

\section*{Acknowledgment}
\label{sec: 5}

I would like to thank all the nice people, who were the first readers of a preliminary version of this paper, for encouraging me that this point of view on 'complex numbers' is worth to communicate to the wider public.

\section*{Trying to publish}

I sent this article to AJP but it was rejected by the editor without referee's opinion.
Next I sent it to EJP. It was rejected but this time the referee had given a comprehensive report.
I believe that he/she takes responsibility for his/her opinion, therefore I cite the report below.
Moreover, I think that the report could be useful for some readers. 
\newline

REPORT ON The simple complex numbers (ref. EJP/256049/PAP)
\newline
This paper is supposed to present a new geometrical interpretation of complex numbers,
different from the conventional one of points in the complex plane; namely, operators on
vectors. The author claims that this is a simpler and more natural interpretation, better
suited for applications in physics and geometry.

However, it is surprising that it is claimed that this approach is new! It is as old as Wessel’s
work (1797) and Hamilton’s struggle towards the conception of hypercomplex numbers
- quaternions -, in the period 1830-1843. See for instance the quotations from Wessel in A.
P. Wills, Vector Analysis with an introduction to tensor analysis, Dover, New York 1958,
pp.xiv-xv; M.J. Crowe A history of vector analysis, Dover, New York 1985, pp7-8; see also
Hamilton’s long preface to his Lectures on Quaternions
in The mathematical papers of Sir William Rowan Hamilton, vol.III, Cambridge University
Press 1967, especially pp. 135-136 and §16, p.123 for the interpretation of $i$ as a rotation
and $i^2 = -1$ as a composition of rotations. Additionally, see Felix Klein’s Development of
Mathematics in the 19th century, Mathematical Science Press, Brookline (MA), 1979 (originally
published in 1928), pp.171-172; L. Sin`egre “Les quaternions et le mouvement du solide
autour d’un point fixe chez Hamilton”, Revue d’histoire des math`ematiques 1, 1995, 83-109,
pp.86-87 for Hamilton’s views concerning the geometric and physical interpretation of operations
with (complex) numbers. Finally, see T. Koetsier “Explanation in the historiography
of mathematics: the case of Hamilton’s quaternions”, Studies in history and philosophy of
science, 26 (no4), 1995, 593-616, pp. 600, 603 for the interpretation of real numbers as
operators. For didactical purposes, this interpretation has been presented in several places,
see e.g. C. Tzanakis “Rotations, complex numbers and quaternions”, Inter. Journal of
Mathematics Education in Science and Technology 26 (no1), 1995, 45-60.

Most of the text, section 2 (9 out of 17 pages) concerns a mathematical subject, namely,
the geometric interpretation of complex numbers as operators. More specifically, section
1 consists of some generalities about the different views of mathematicians and physicists.
Section 2, which is the main part of the paper, concerns elementary results, well known to
the average mathematician:

(a) pp.4-5: That (real) numbers can be considered as multiplicative operators: mathematically
expressed, this means that there is an isomorphism betwee real numbers as a
multiplication group and the group of automorphisms of R onto itself
$f_{a}(x) = ax$

(b) pp.6-7: The extension of this fact to include scalar multiplication of 2D-vectors, i.e.
that real numbers can be seen as multiplication operators on vectors, that is, that R, as an
algebraic field is used to define R2, R3 etc as vector spaces over this field.

(c) pp.7-10: The extension of the idea in (b) from R to C, so that considering R2 as
a vector space over C, multiplication of a vector by a complex number z amounts to the
composition of a rotation (by $arg(z)$) and a dilatation (by $|z|$). In other words, considering
the dual nature of complex numbers as a multiplication group (in the sense of (a) above)
and as a vector space (isomorphic to) R2, the product z · w amounts to the action on w of
the composition of the rotation and the dilatation induced by z. This is known to anyone
who has studied de Moivre’s trigonometric form of complex numbers and their geometric
interpretation and can be found in standard textbooks like K. Knopp Elements of the theory
of functions, Dover 1952, §16.

(d) p.11: The motivation for the definition of the complex exponential function can be
found in standard calculus textbooks, e.g. T. Apostol Calculus, Xerox College Publishing,
1967, vol.I, §9.7.
Section 3 gives some quite elementary applications of this well known interpretation of
C. I consider the last part of section 3 (pp.16-17) and section 4 as a set of remarks, which
do not add anything essential to the main text.
In view of these comments, I do not recommend the publication of this paper to EJP.

\section*{My answer to EJP editor}

Well, it is certainly your decision. I do not want to argue with the referee.

But I hope you see the difference between my simple and clear explanation of the "complex numbers" and the mathematical bla bla bla of the referee. Using my approach you can explain the idea of "complex numbers" even to pre-school pupils. I am sure that referee's language would be quite unpalatable for them. That's the difference. I stress the visualisation and "motion" of the "complex numbers". I show that they are so simple as waving with a broomstick. For the referee more important is a formal structure of the set.

Moreover, the textbooks are full of such unnecessary or at least second-rate notions as "imaginary axis", "complex plane", and so on. It is the main difference between my interpretation and the standard one.

Maybe my approach is not as new as I thought, maybe it is a re-discovery of some ideas of Wessel, Grassmann, Hamilton. Nevertheless, I completely do not agree with the referee that you can find it in a standard textbook.
\\
Yours sincerely  \\
                                  Jaroslaw Zalesny

\end{document}